\documentclass[aps,prx,preprint,a4paper]{revtex4-2}

\usepackage[utf8]{inputenc}

\usepackage{graphicx}
\usepackage{float}

\usepackage{amsmath}
\usepackage{physics}
\usepackage{hyperref}
\usepackage{xcolor}

\begin{document}

\title{A Ballistic Two-Dimensional Lateral Heterojunction \\Bipolar Transistor}

\author{Leonardo Lucchesi}
\email[]{leonardo.lucchesi1@phd.unipi.it}
\affiliation{Department of Physics ''Enrico Fermi'', University of Pisa, Pisa}
\affiliation{Department of Information Engineering, University of Pisa, Pisa}
\author{Gaetano Calogero}
\affiliation{Department of Information Engineering, University of Pisa, Pisa}
\author{Gianluca Fiori}
\affiliation{Department of Information Engineering, University of Pisa, Pisa}
\author{Giuseppe Iannaccone}
\email[]{giuseppe.iannaccone@unipi.it}
\affiliation{Department of Information Engineering, University of Pisa, Pisa}

\begin{abstract}
We propose and investigate the intrinsically thinnest transistor concept: a monolayer ballistic heterojunction bipolar transistor based on a lateral heterostructure of transition metal dichalcogenides. The device is intrinsically thinner than a Field Effect Transistor because it does not need a top or bottom gate, since transport is controlled by the electrochemical potential of the base electrode. As typical of bipolar transistors, the collector current undergoes a tenfold increase for each 60 mV increase of the base voltage over several orders of magnitude at room temperature, without sophisticated optimization of the electrostatics. We present a detailed investigation based on self-consistent simulations of electrostatics and quantum transport for both electron and holes of a pnp device using MoS$_2$ for the 10-nm base and WSe$_2$ for emitter and collector. Our  three-terminal device simulations confirm the working principle and a large current modulation I$_\text{ON}$/I$_\text{OFF}\sim 10^8$ for $\Delta V_{\rm EB}=0.5$~V. Assuming ballistic transport, we are able to achieve a current gain $\beta\sim$~10$^4$ over several orders of magnitude of collector current and a cutoff frequency up to the THz range. Exploration of the rich world of bipolar nanoscale device concepts in 2D materials is promising for their potential applications in electronics and optoelectronics.
\end{abstract}

\maketitle

\section*{Introduction}
The bipolar junction transistor (BJT) has been the first semiconductor transistor manufactured in volume~\cite{shockley1948} and for thirty years the workhorse of semiconductor electronics, before being taken over by the metal-oxide-semiconductor field-effect transistor (MOSFET). Still, as of today, the Heterojuction Bipolar Transistor (HBT) is the fastest transistor~\cite{quan2018}, and is the device of use in applications where high power and very high frequency are required, such as telecommunication stations and satellite communications. In addition, BJTs are largely used in ubiquitous building blocks of integrated circuits, such as bandgap voltage references and temperature sensors.\\
Indeed, device physicists and engineers are much more familiar with MOSFETs than BJTs, and the recent explosion of interest for electron devices based on 2D materials has been mainly focused on MOSFETs~\cite{Fiori2014, Iannaccone2018}, for the possibility of enabling an extension, or even an acceleration, of the so-called Moore's law, i.e., the exponential increase of the number of transistors in an integrated circuit as a function of time. This possibility is predicated on the fact that 2D materials can provide an extremely thin layer with a relatively high mobility, thereby enabling scaling of channel width and length while preserving a good electrostatic behavior and low delay times.\\ 
\begin{figure*}
\centering
\includegraphics[width=0.9\textwidth]{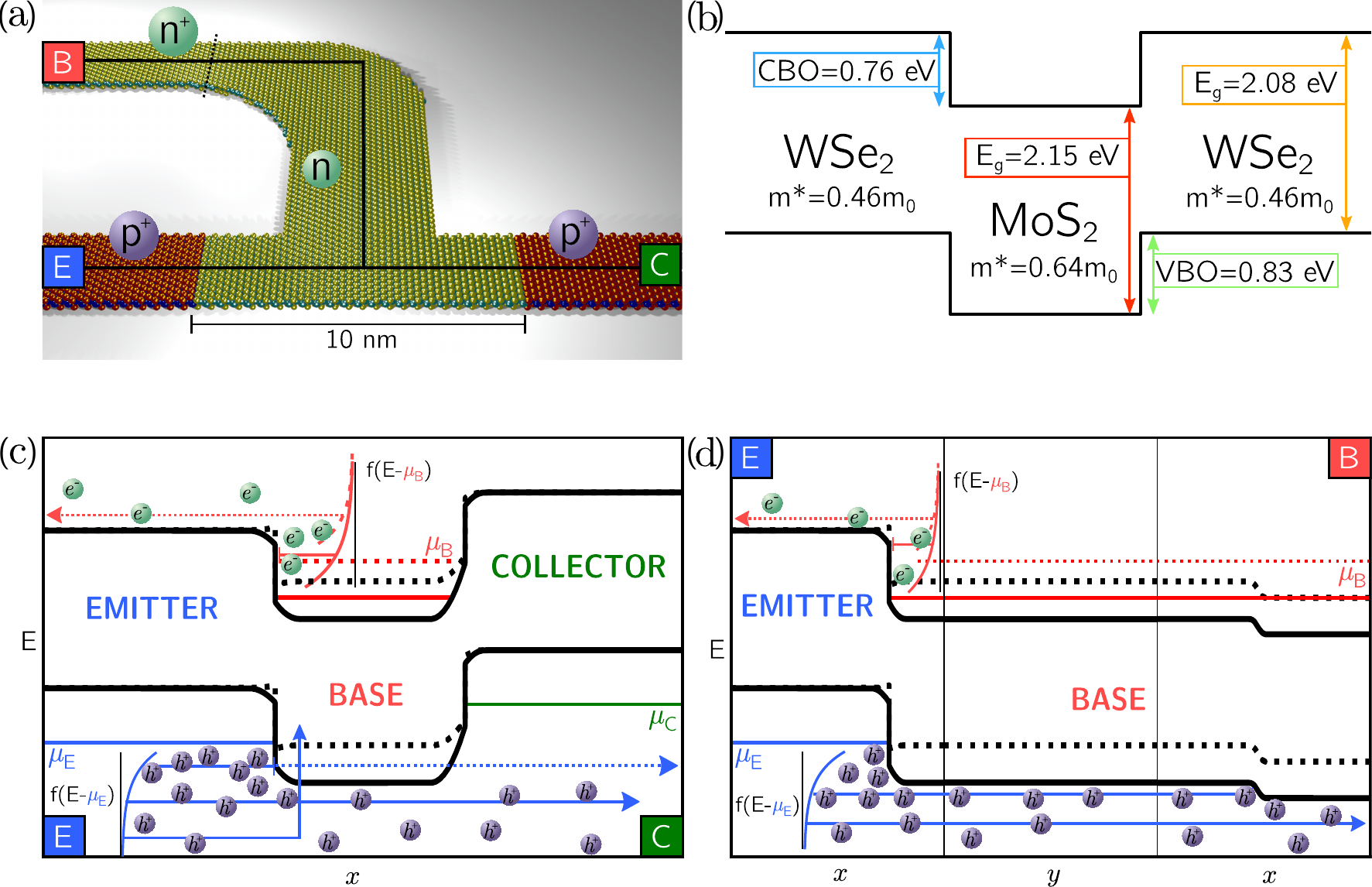}
\caption{Device layout and working principle. (a) Artistic illustration of the device. We indicate materials doping and the Emitter-Collector(EC) and Emitter-Base(EB) paths along which we show most physical quantities. Leads are represented with letters: Emitter, Base and Collector. (b) Band alignment of materials (App.~\ref{appa}). Working principle is represented in (c) for the EC path, and in (d) for the EB path. Black lines are band edges (solid $\mu_B=1.90$ eV, dotted $\mu_B=2.38$ eV). Blue/green lines are emitter/collector chemical potentials ($\mu_E=0.0$ eV, $\mu_C=0.5$ eV), and red solid/dotted lines are the base chemical potential for $\mu_B=1.90/2.38$ eV. Arrows and spheres represent carrier flows, the upward arrow in the middle plot is the emitter to base hole flow. For $\mu_B=1.90$ eV, the base barrier blocks hole flow from E to B (OFF state). When $\mu_B$ is raised to $2.38$ eV, the barrier rises, and more holes flow (ON state). However, electron flow from B to E (dotted red arrow in both plots) and hole flow from E to B are also increased. A more detailed discussion can be found in text.
 }
\label{figure1}
\end{figure*}
While BJTs and MOSFETs are very similar devices in terms of potential-barrier-controlled transport, BJTs have the advantage of the collector current increasing by a factor 10 for every 60 mV of increase of the base voltage for several decades, at room temperature, up to very high current density. The drain current of a MOSFET has a similar exponential dependence on the gate voltage only at very low current, in the so-called "subthreshold region", and shows a slower exponential. The reason for this difference is that in both devices the main current is controlled by modulating the barrier between the central region (the base for the BJT and the channel for the MOSFET) and the emitting electrode (the "emitter" in the BJT and the "source" in the MOSFET). The advantage of the BJT, in this case, is simply that the base region is in direct contact with a metal electrode, therefore a change of the voltage applied to the base electrode is directly transferred to the electrostatic potential in the base, whereas in the case of the MOSFET the channel region is only capacitively coupled to the gate electrode through the gate dielectric layer, and there is a voltage partition between the potential on the gate and the potential in the channel.We highlight the negligible gate current as the big advantage of the MOSFETs, whereas the BJT has a non negligible base current, that is a factor $\beta$ smaller than the collector current (where the current gain $\beta$ is in the range $10$-$10^3$), and that typically increases the power consumption of circuits based on BJTs. At the nanoscale and at very low current density, BJTs can again be promising as compared to MOSFETs, because a subthreshold swing close to 60 mV/decade in BJTs is obtainable without the use of extremely thin dielectric layers and sophisticated electrostatics engineering, and because source-to-drain leakage currents can become comparable to base currents, if $\beta$ is sufficiently high. \\
Among the classes of 2D materials, TMDCs are one of the most promising for transistor use~\cite{Radisavljevic2011,Fang2012,Jariwala2014,Manzeli2017}, offering a wide range of properties and supporting relatively inexpensive, stable and scalable fabrication techniques. Among those, one-step Chemical Vapor Deposition (CVD) growth is establishing a quality standard for devices~\cite{Manzeli2017,Cai2018}, but the realization of TMDC lateral p-n junctions is still a complex issue~\cite{Li2015}. Two contributions in this direction stand out: a two-step CVD growth by Li et al.~\cite{Li2015} that allows the growth of atomically thin WSe$_2$-MoS$_2$ p-n junctions, and a morphological method by Han et al.~\cite{Han2017} for growing narrow MoS$_2$ channels embedded in WSe$_2$. The type-II heterostructure shown in Fig.~\ref{figure1}(b) is obtained in both cases, making these methods suitable for the production of Heterostructure Bipolar Junction Transistors (HJT) as demonstrated by Lin et al.~\cite{Lin2017}.\\
In this paper, we propose the concept of a ballistic lateral heterojunction bipolar transistor based on Transition Metal DiChalchogenides (TMDC), and assess its potential in electronics applications with self-consistent quantum transport and electrostatics simulations using the Non-Equilibrium Green's Function (NEGF) formalism~\cite{Lake1997,Datta2005}. 
We highlight that, despite the fact that BJTs have received little attention by the vibrant global 2D materials research community, some experimental demonstrations have recently appeared in the literature, based both on lateral~\cite{Agnihotri2016,Lin2017} and vertical~\cite{Liu2019,Su2020} operation. The main challenges in the fabrication of 2D BJTs reside in effectively doping the regions and in obtaining high quality heterojunctions. From a computational point of view, the main challenge is the modelling of two-carrier flows in a far-from-equilibrium multilead setup, that is a notoriously challenging problem and has been added as a capability to the NanoTCAD ViDES package~\cite{NanoTCAD}, the software used for all simulations in this work. 
\section*{Methods}
\subsection*{Model}
In order to describe our system, we use a nearest-neighbour effective mass tight-binding model~\cite{Agarwal2017}. In this model, materials are described by a tight-binding Hamiltonian whose bandstructure has the same bandgap, electron affinity, and band curvature (effective mass) as the original material. Transport is assumed to be completely ballistic (see App.~\ref{appb}). The system is solved by using an NEGF-Poisson iterative self-consistent procedure, where we use NEGF to compute currents between leads and the charge ballistically injected from the leads into the device region, solve a Poisson problem for that charge and use the resulting potential as an input to the next NEGF step. Simulation of device operation requires proper handling of two main issues: the presence of three leads in a far-from-equilibrium condition common in real nanoscale devices and the need of a self-consistent description of transport and electrostatics. 
As we are considering TMDCs, we can use an hexagonal lattice with lattice parameter $d$ and two effective atoms in the primitive cell, each with a single energy level. The Hamiltonian reads
\begin{equation}
\begin{bmatrix}
E_0 & t f(\vec{k})\\
t f(\vec{k}) & E_1
\end{bmatrix}
\end{equation}
where $f(\vec{k})=(1+e^{-i\vec{k}\cdot\vec{a}})$
The bands generated by this Hamiltonian have a direct gap $E_{\rm gap}=E_0-E_1$ in the $K$-point. Close to $K$, the dispersion relation can be described as free, with an effective mass that is related to the hopping parameter $t$ by
\begin{equation}
m^*=\frac{\hbar^2}{q}\frac{2 \Delta}{9 d^2 |t|^2},
\end{equation}
where $d$ is the interatomic distance and $q$ is the electron charge. In our nearest-neighbor model, the effective mass results to be the same for the valence and the conduction band, restricting the validity of this Hamiltonian to those materials whose electron and hole effective masses are similar. Both WSe$_2$ and MoS$_2$ fall into this category (see App.~\ref{appa}). Heterojunctions have been described via a linear interpolation of all Hamiltonian parameters in their proximity.

\section*{Results}
\subsection*{Working principle}
\begin{figure*}
\centering
\includegraphics[width=0.9\textwidth]{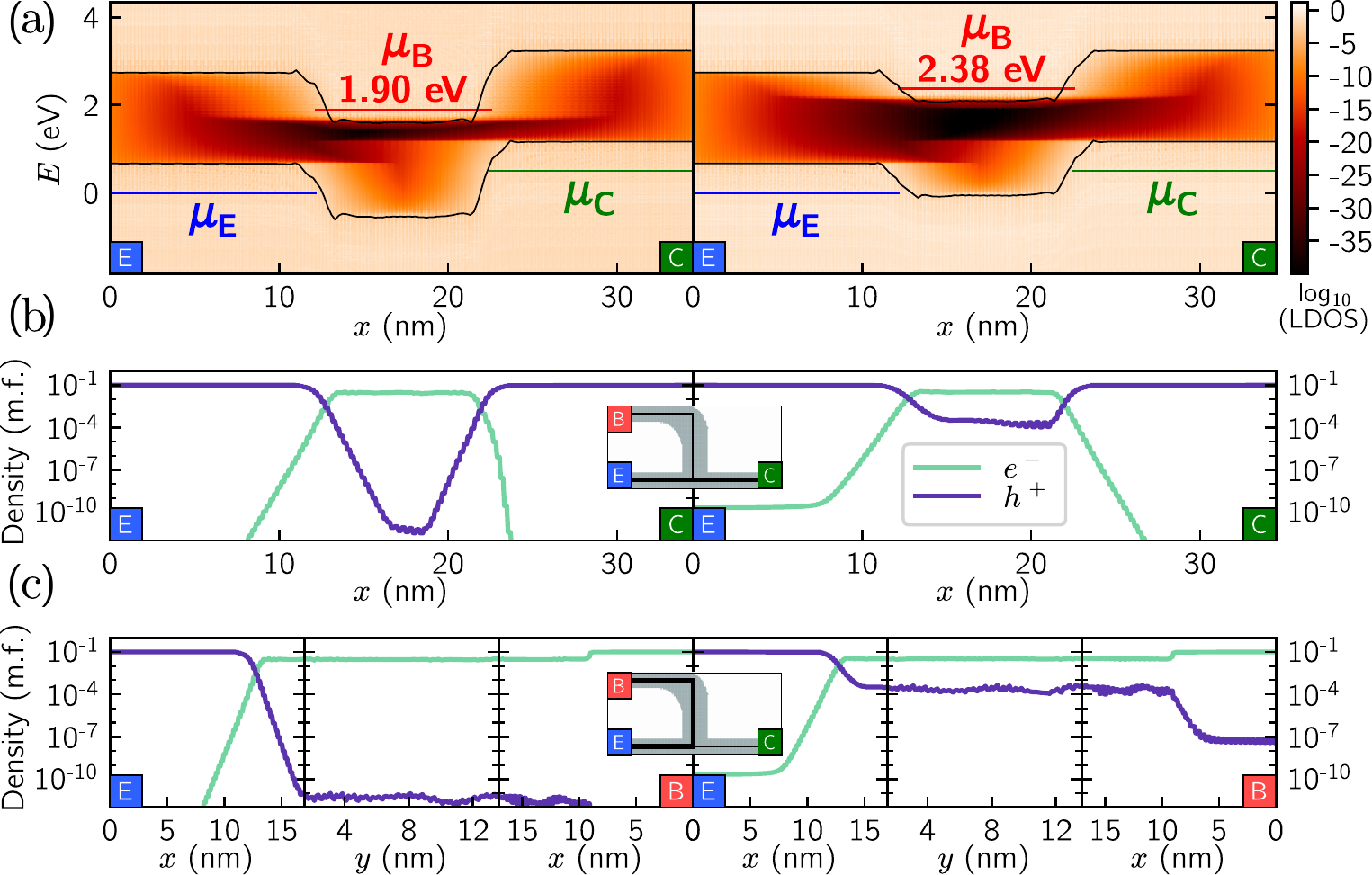}
\caption{Demonstration of working principle.(a): Local Density Of States (LDOS) along the EC path for $\mu_B=1.90/2.38$ eV. Black lines represent band edges, colored straight lines are lead chemical potentials. Change in LDOS following change $\Delta\mu_B$ confirms working principle: main difference is a $\Delta\mu_B$ rigid shift in the bands (no gate present). (b) and (c): Injected carrier concentrations (molar fraction) respectively for EC path and EB path, both shown for $\mu_B=1.90/2.38$ eV . Insets are path references. Charge neutrality is confirmed in the whole device, central region included (non-trivial), as injected carrier concentration is equal and opposite to doping. Concentrations lower than $10^{-12}$ are neglectable.}
\label{figure2}
\end{figure*}

We consider a p-n-p double heterojunction bipolar transistor (DHBT) with 2D materials, with an MoS$_2$ base, and WSe$_2$ emitter and collector. We consider a nanoribbon channel and the possibility of heavily doping the 2D regions, so that the device structure under investigation is illustrated in Fig.~\ref{figure1}(a). Let us stress that transistors work correctly even if there is no lateral confinement, and that effective doping is required for a device practically usable in circuits. Therefore, in order to assess the potential of 2D BJTs, we need to assume that technological challenges in doping of 2D materials on a large scale can be solved, as has been demonstrated in laboratory conditions~\cite{Fang2012,Radisavljevic2011}.\\ 
Device operation is represented in Fig.~\ref{figure1}(c)-(d) through ideal band edge profiles, respectively along the emitter-to-collector path and along the emitter-to-base path. A thermionic hole current flowing between the emitter and collector WSe$_2$ regions is controlled by modulating a barrier in the central MoS$_2$ region (base). This barrier is modulated by changing the electrochemical potential of the base lead, i.e. injecting electrons in the base region. This modulation causes a shift in the electrostatic potential in the base through the unbalance between the charge injected from the leads and the fixed charge due to materials doping. We can find an easy interpretation for this shift by inspecting the bulk of each region, i.e. more than a screening length away from heterojunctions and their depletion layers. In the bulk, local equilibrium should ensure a vanishing electric field. Otherwise, carriers would flow and neutralize it. This corresponds to a constant electrostatic potential, only obtainable with charge neutrality. The main currents components are represented by hole flow from emitter to collector over the barrier ($I_C$), electron flow from base to emitter ($I_{BE}$) and hole flow from emitter to base ($I_{EB}$), as shown in Fig.~\ref{figure1}(c)-(d). The base current $I_B$ is obtained as $I_B = I_{BE}+I_{EB}$. In order to keep the current gain $\beta=I_C/I_B$ as high as possible, we need to keep the carrier flow to and from the base as small as possible. While $I_{BE}$ can be controlled independently via the difference between the emitter conduction band edge and $\mu_B$ (e.g. lowering base doping), $I_{EB}$ and $I_C$ are both controlled by the difference between the emitter chemical potential $\mu_E$ and the base valence band edge. This issue can be solved by introducing a barrier for holes in the base lead, here obtained by incrementing donor doping in the base lead (Fig.~\ref{figure2}(c)). The flat potential profile in the base and the working principle are demonstrated in Fig.~\ref{figure2}(a), where we show a section of the Local Density Of States (LDOS) along a path from emitter to collector for $\mu_B=1.90$ eV and $2.38$ eV. As expected, we find that the LDOS away from the heterojunctions corresponds to the bulk LDOS shifted by a flat electrostatic potential. The carrier injection depicted in Fig.~\ref{figure2}(b)-(c) also confirms the working principle, as we can see that the fixed charge has been effectively neutralized by the injected carriers in all leads, ensuring the charge neutrality condition for both the ON and the OFF states. 
\subsection*{Device characteristics}
\begin{figure*}
\centering
\includegraphics[width=0.9\textwidth]{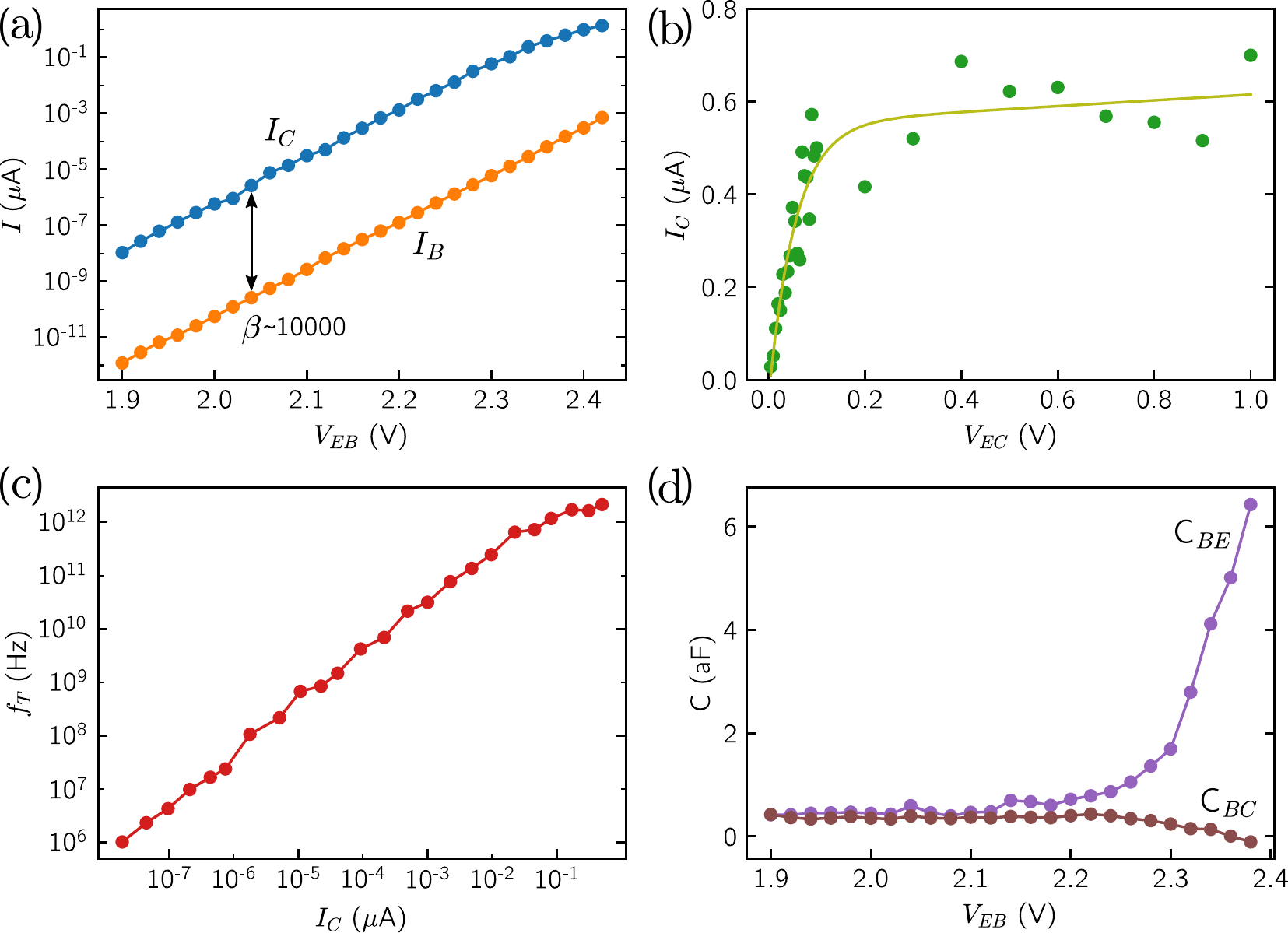}
\caption{Performance assessment. (a) IV plot showing dependence of collector $I_C$ and base $I_B$ currents on $V_{EB}\equiv q^{-1}(\mu_B-\mu_E)$ for $V_{EC}\equiv q^{-1}(\mu_C-\mu_E)$ fixed to $V_{EC}=0.5$ V. Both currents show exponential behavior with $\sim 60$ mV/decade slope. The almost constant current gain $\beta=I_C/I_B$ is easily inferrable to be $\sim10000$, dropping to $\sim6000$ deep in the ON state because of saturation.(b) Dependence of collector current $I_C$ on $V_{EC}$ for $V_{EB}=2.38$ V (ON state). The Ebers-Moll model fits our results with a reasonable mismatch. (c) The cutoff frequency $f_T$ as a function of the collector current $I_C$. (d) Capacitances' $C_{BE}$ and $C_{BC}$ dependence on $V_{EB}$. Both are almost constant for most of our voltage span.}
\label{figure3}
\end{figure*}

In Fig.~\ref{figure3}(a) we plot the collector current $I_C$ and the base current $I_B$ as a function of  $V_{EB}\equiv q^{-1}(\mu_B-\mu_E)$. Both curves behave exponentially with a current swing of $\sim 60$ mV/decade, typical for very well controlled thermionic currents~\cite{Sze2006}. In the case of the MOSFET we would have called this quantity "subthreshold swing", but in the case of BJT it would not be appropriate, since the exponential behavior extends for several orders of magnitude up to very high current density. This implies that the electrostatic potential in the base closely follows the electrochemical potential  $\Delta \phi_B \simeq \Delta \mu_B$ i.e., as the potential is controlled just by modifying $\mu_B$, i.e. the carrier injection from the base lead. The very steep exponential dependence of $I_C$ on $V_{EB}$ enables a large current modulation $(I_{ON}/I_{OFF}) \simeq10^8$ for $\Delta V_{EB}=0.5$ V, higher than what is typically achievable with nanoscale MOSFETs. As in common BJTs, increased control comes with the price of a finite base current $I_B$. The current gain $\beta=I_C/I_B$ is highlighted in Fig.~\ref{figure3}(a), reaching the value of $\beta \simeq 10^4$. For larger $V_{BE}$, $\beta$ drops to $\sim 6000$. This is due to saturation, as we can also see from the LDOS in Fig.~\ref{figure2}(a) with the base barrier being above the emitter chemical potential.
This large value for $\beta$ is mainly due to the presence of the barrier for holes in the base lead, shown in Fig.~\ref{figure1}(d). Without that barrier, the best $\beta$ obtainable would be approximately $\sim 5$, close to what was experimentally observed in~\cite{Lin2017}. We implemented this barrier by increasing doping, but in principle it could be created in different ways, e.g. via a Schottky junction or electrostatic doping with a gate electrode. The plot in Fig.~\ref{figure3}(b) represents the dependence of the collector current $I_C$ on the emitter-collector bias $V_{EC}$ in the ON state. Here, the noise deriving from the suboptimal convergence is more evident because of the linear scale. Despite the noise, the simulation results can be fitted with an Ebers-Moll model, confirming that our simulation correctly reproduces the essential physics of this device. The cut-off frequency $f_T$ is computed quasi-statically as 
\begin{equation*}
f_T = \frac{ \frac{\partial I_C}{\partial V_{EB}} }{ 2 \pi \frac{\partial Q}{\partial V_{EB}}}
\end{equation*}
where Q is the total mobile charge in the device, and is shown in Fig.~\ref{figure3}(c) again showing signs of convergence-related noise. Given our assumptions of ballistic transport and high doping in the base, the cut-off frequency steeply increases with the collector current of several orders of magnitude up to the THz regime, and never reaches the high-injection regime. Both the high $\beta$ and the cut-off frequency results are very promising and are connected to the extremely short base of the BJT, to the high doping, and to the assumption of ballistic transport in the base (see App.~\ref{appb}). This means that they are optimistic if compared to present-day fabrication capability, but they indicate potential for applications.
The differential capacitances $C_{BE}=\pdv*{Q_{E}}{\mu_B}$ and $C_{BC}=\pdv*{Q_{C}}{\mu_B}$, where $Q_E\,(Q_C)$ is the total charge injected by E(C), are plotted in Fig.\ref{figure3}(d) as a function of $V_{EB}$. 
\section*{Conclusion}
We have proposed and investigated the device concept for the intrinsically thinnest transistor: a nanoscale 2D double heterojunction bipolar transistor using a lateral II-type heterostructure of WSe$_2$ and MoS$_2$. 
We have shown that this device concept preserves many of the well-known beautiful features of traditional bipolar transistors, such a tenfold increase of the current for an increase of 60 mV of the base-emitter voltage (a "current swing" of 60 mV/decade) over several orders of magnitude of collector current, which is hard to obtain in the case of nanoscale MOSFETs and would require sophisticate multi-gate and electrostatic engineering.
We have also shown that in principle high current gain of few thousands is achievable in the ballistic transport assumption if the base lead is properly engineered, even at low current bias, much higher than obtained in most experiments~\cite{Lin2017,Liu2019,Su2020} and comparable to what has been obtained in the best case~\cite{Agnihotri2016}.
The proposed device is intrinsically promising for high frequency applications, in terms of cut-off frequency and beta, and also for high performance digital applications, when MOSFET leakage currents are comparable with the base current of the BJT. In order for this potential to be actually achieved, several technology improvements have to be achieved, in particular concerning the possibility of doping TMDCs at large molar fractions, and of fabricating high quality heterojunctions and short base regions.

\subsection*{Acknowledgements}
The work has been partially supported by the European Commission through the h2020 FET project QUEFORMAL (contract number n. 829035) and by the Italian Ministry of University and Research through the PRIN project FIVE2D (contract number 2017SRYEJH\_001)
\appendix
\section{Simulation setup}\label{appa}
Within our model, any material can be described by three parameters: conduction band edge $E_{CB}$, valence band edge $E_{VB}$ and a single effective mass $m^*$. Our choice of materials and setup has been inspired by recent work on the creation of lateral heterostructures~\cite{Li2015,Han2017}. Our device is built to be a T-shaped heterostructure, where the central region is made of MoS$_2$ and the lateral regions are made of WSe$_2$. We used the band alignment from~\cite{Chiu2015} and the hole effective masses from~\cite{Chang2014}, setting $E_{\rm gap,WSe_2}=2.07$ eV, $E_{\rm gap,MoS_2}=2.15$ eV, CBO$=0.76$ eV (Conduction Band Offset, MoS$_2$ lower), $m^*_{\rm WSe_2}=0.46\,m_0$ and $m^*_{\rm MoS_2}=0.64\,m_0$, as shown in Fig.~\ref{figure1}(b). The approximation of taking the hole effective mass for both carriers is reasonable for the materials considered in this work, since for $m^*_{e,\rm WSe_2}=0.35\,m_0$ and $m^*_{e,\rm MoS_2}=0.56\,m_0$. The use of a single effective mass comes from the nearest-neighbor hopping model. A next-to-nearest neighbor model would allow us to have different effective masses for electrons and holes, but it would force us to double the dimension of the subsystems used for the computation (see below), leading to an increase in computational time of $2^{2.7}\sim6.5$ times. Since the source-drain current is carried by holes, we believe that this approximation does not qualitatively affect our results.\\
We computed the system Green's function via the Recursive Green's Function (RGF) algorithm~\cite{Thouless1981,Lake1997} implemented in NanoTCAD ViDES. RGF requires the division of the system in parts called ''slices'' along the transport direction, interacting only between nearest-neighbors. Every matrix quantity such as Green's function or self-energies is divided in blocks corresponding to slices. 
While this procedure is best suited for two leads attached to entire single slices, we implemented the third lead through the Meir-Wingreen formula for transmission~\cite{Meir1992} and the spectral function
\begin{align*}
T_{I,J}&=\text{Tr}\left( G_{ij}^\dagger \Gamma_I G_{ij} \Gamma_J \right)\\
A_{I\alpha}&=G_{\alpha i} \Gamma_I G_{i\alpha}^\dagger
\end{align*}
where $i,j$ are the device sites belonging to leads $I,J$, $\alpha$ is a site belonging to the device, $G_{ij}^{(\dagger)}$ are the Green's function blocks corresponding to the sets of atoms $i,j$ and $\Gamma_I=i(\Sigma_I - \Sigma_I^\dagger)$ is the broadening matrix corresponding to the lead self-energy $\Sigma_I$. The implementation runs through the broadening matrices, used as projectors on the set of sites belonging to the corresponding lead. If the non-zero parts of the broadening matrices of two leads do not intersect with each other, we may have then two leads attached to the same slice of the device. Further details can be found in the SI~\cite{SI}. \\ The doping molar fraction is $10^{-1}$ for n$^+$ and p$^+$ and $3\times10^{-2}$ for n, which are higher than experimentally feasible, but required for numerical stability of the computation. Indeed, the fixed charge $\rho_{fixed}$ makes the solution of the Poisson equation
\begin{equation}
\nabla\cdot(\epsilon\,\nabla\phi)=4\pi(\rho_{free}+\rho_{fixed})
\end{equation}
stable towards variations in the free charge $\rho_{free}.$ We believe this assumption of high doping does not affect the conclusions of this work, as only the presence of doping and the ratios of dopant densities between adjacent regions are important for the device operation described above.

\section{Device optimization and ballistic approximation}\label{appb}
We can summarize the device geometry as a T shaped MoS$_2$ structure, with the horizontal part being a $\sim10\times3$ nm armchair nanoribbon and the vertical part being a $\sim5$ nm wide zigzag nanoribbon. Two WSe$_2$ armchair nanoribbons extend the horizontal part, forming two lateral heterostructures. The vertical part forms a curve to have an horizontal end, because the Recursive Green's Function algorithm requires leads oriented in the transport direction, as self energies can only connect neighboring slices~\cite{SI}. The presence of a curve represents just a numerical complication, as the general device mechanism only requires a generic source injecting carriers in the central region. We consider the shortest possible base that avoids punch-through~\cite{Sze2006}, that in this case is 10 nm~\cite{SI}. This makes it possible for us to assume ballistic transport in the base region, considering that the mean free path in MoS2 is 8 nm~\cite{Radisavljevic2011,Sze2006} and making an approximation that is justiﬁed in the context of a device concept investigation, since our results do not critically depend on coherence. In this regime, a large part of carriers ($\sim$50\%) would propagate freely through the base behaving as we described. The other part would undergo phonon scattering events while propagating through the base, mostly losing phase coherence, but retaining most of the initial momentum. Therefore, we believe that a limited amount of scattering would not critically aﬀect our ﬁgures of merit, which depend mainly on charge transport, not on phase coherence.\\


\end{document}